\RequirePackage{fix-cm}
\documentclass[fleqn]{svjour3}          
\smartqed  
\usepackage{graphicx}
\usepackage{mathptmx}      
\usepackage{amssymb}
\usepackage{amsfonts}
\usepackage[english]{babel}
\usepackage{euscript}
\usepackage{bbm}
\usepackage{xfrac}
\usepackage{color}
\newcommand{\fl}{\hspace*{-0pt}}

\newcommand{\eqref}[1]{(\ref{#1})}

\newcounter{example_counter}
\newcounter{remark_counter}
\begin{document}

\title{Integrability structures of the generalized Hunter--Saxton equation}

\titlerunning{Integrability structures of the generalized Hunter--Saxton equation}

\author{Oleg I. Morozov}

\institute{Oleg I. Morozov
        \at
        Faculty of Applied Mathematics,
        AGH University of Science and Technology,
        \\
        Al. Mickiewicza 30,  Cracow 30-059, Poland
        \\
        \email{oimorozov@gmail.com}
        }

\date{Received: date / Accepted: date}

\maketitle

\begin{abstract}

We consider integrability structures of the generalized Hunter--Saxton equation. In particular, we obtain the
Lax representation with nonremovable spectral parameter, find local recursion operators  for symmetries and
cosymmetries, generate an infinite-dimensional  Lie  algebra of higher symmetries, and prove existence of
infinite number of cosymmetries of higher order. Further, we  give an example of employing the higher order
symmetry to constructing exact globally defined solutions for the generalized Hunter--Saxton equation.



\end{abstract}

\section{Introduction}

The Hunter--Saxton equation
\begin{equation}
u_{tx} = u\,u_{xx}+\textstyle{\frac{1}{2}}\,u_x^2
\label{HS_eq}
\end{equation}
was introduced in \cite{HunterSaxton1991} to describe the nonlinear instability of the director field in
the nematic liquid crystal and then has been a subject of thorough investigation. As it was shown in
\cite{HunterSaxton1991}, equation \eqref{HS_eq} admits a Lagrangian formulation with Lagrangian
$L = (u_t-u\,u_x)\,u_x$. In \cite{HunterZheng1994} a bi-Hamiltonian structure, a Lax representation, a nonlocal
recursion operator, and a series of conservation laws  have been found. A tri-Hamiltonian formulation for
\eqref{HS_eq} was proposed in \cite{OlverRosenau1996}. Inverse scattering solutions for \eqref{HS_eq} were
constructed in \cite{BealsSattingerSzmigielski2001}. In \cite{KhesinMisiolek2003} it has been proven that
equation \eqref{HS_eq} can be understood as a geodesic equation associated to a right-invariant metric on an
appropriate homogeneous space related to the Virasoro group. The pseudo-spherical formulation for equation
\eqref{HS_eq} and  quadratic pseudopotentials were proposed and used to find nonlocal symmetries and conservation
laws in \cite{Reyes2002}. In \cite{GorkaReyes2012}, the nonlocal symmetries were used to construct exact
solutions and a nonlocal recursion operator for \eqref{HS_eq}. Nonlocal recursion operators,  a fourth order
local recursion operator, series of higher symmetries  and conservation laws for equation \eqref{HS_eq} have
been constructed in \cite{Wang2010}, see also \cite{TianLiu2016}.

The  further discussion of the physical interpretation of equation \eqref{HS_eq} can be found in
\cite{BressanConstantin2005}.

In this paper we consider the generalization
\begin{equation}
u_{tx} = u\,u_{xx}+\beta\,u_x^2, \qquad \beta \neq 0,
\label{gHS_alpha}
\end{equation}
of the Hunter--Saxton equation \eqref{HS_eq}. This equation  with $\beta \neq \frac{1}{2}$ has applications in
geometry of Einstein--Weil structures \cite{Tod2000,Dryuma2001}, and in hydrodynamics \cite{Golovin2004}.
In \cite{Calogero1984,Pavlov2001} a nonlocal transformation was used to construct a general solution for
 \eqref{gHS_alpha}. In \cite{Morozov2005} we have shown that equation   \eqref{gHS_alpha} is linearizable via the
contact transformation
$(t,x,u,u_t,u_x) \mapsto (\tilde{t},\tilde{x},\tilde{u},\tilde{u}_{\tilde{t}},\tilde{u}_{\tilde{x}})$
given by the formulae
\begin{equation}
\left\{
\begin{array}{lcl}
t &=& \beta^{-1}\,\tilde{t},
\\
x &=& -(\tilde{t}+\tilde{x})^{\frac{\beta-1}{\beta}}\left(\beta\,(\tilde{t}
+\tilde{x})\,\tilde{u}_{\tilde{x}}- \tilde{u}\right),
\\
u &=&(\tilde{t}+\tilde{x})^{-\frac{1}{\beta}}\left(\beta\,(\tilde{t}+\tilde{x})\,\tilde{u}_{\tilde{x}}
+(\beta-1)\,\tilde{u}\right),
\\
u_t &=& \beta^2\,(\tilde{t}+\tilde{x})^{-\frac{1}{\beta}}\left(\tilde{u}_{\tilde{t}}
-\tilde{u}_{\tilde{x}} \right),
\\
u_x &=& - (\tilde{t}+\tilde{x})^{-1}.
\end{array}
\right.
\label{HS_EP_transform}
\end{equation}
This transformation maps \eqref{gHS_alpha} to the Euler--Poisson equation
\begin{equation}
\tilde{u}_{\tilde{t}\tilde{x}} = \frac{1}{\beta\,(\tilde{t}+\tilde{x})}\,\tilde{u}_{\tilde{t}}
+ \frac{2\,(1-\beta)}{\beta\,(\tilde{t}+\tilde{x})}\,\tilde{u}_{\tilde{x}}
- \frac{2\,(1-\beta)}{(\beta\,(\tilde{t}+\tilde{x}))^{2}}\,\tilde{u}.
\label{EP}
\end{equation}
In its turn equation \eqref{EP} in integrable by quadratures via La\-pla\-ce's method, \cite[\S~9.3]{Ovsyannikov}.
The general solution to \eqref{EP} combined with the inverse transformation to \eqref{HS_EP_transform} provides
the parametric formula for the general solution to equation \eqref{gHS_alpha}, see details in \cite{Morozov2005}.
This formula is locally defined and does not give global solutions to \eqref{gHS_alpha}.

In the present paper we study integrability properties of equation \eqref{gHS_alpha}. In Section
\ref{Lax_representation_section} we find the Lax representation for \eqref{gHS_alpha} with arbitrary $\beta$.
We show that this Lax representation includes the non-removable spectral parameter. We study contact symmetries
of this equation in Section \ref{contact_symmetries_section}. We show that the Lie algebra of contact symmetries
of equation \eqref{gHS_alpha} is the semi-direct sum $\mathfrak{s}_4 \ltimes \mathfrak{a}_\infty$ of the
four-dimensional Lie algebra $\mathfrak{s}_4 \cong \mathfrak{gl}_2(\mathbb{R})$ and the infinite-dimensional
Abelian ideal $\mathfrak{a}_\infty$. Then in Section \ref{recursion_operators_section} we apply the approach of
\cite{KrasilshchikKersten1994,KrasilshchikKersten1995,KrasilshchikKersten2000,KrasilshchikVerbovetskyVitolo2017}
to find local and nonlocal recursion operators for symmetries of \eqref{gHS_alpha}. In Section
\ref{higher_symmetries_section} we study the action of local recursion operators to the subalgebra
$\mathfrak{s}_4$. This action generates a Lie subalgebra $\mathfrak{s}_\infty$ of the algebra of higher
symmetries of equation \eqref{gHS_alpha}. We show that $\mathfrak{s}_\infty$ has an interesting structure of the
so-called Lie algebra of matrices of complex size, \cite{Feigin1988,PostHijligenberg1996}. Cosymmetries of
\eqref{gHS_alpha} and recursion operators for cosymmetries are discussed in Section \ref{cosymmetries_section}.
Finally, in Section \ref{invariant_solutions_section} we use a higher symmetry from $\mathfrak{s}_\infty$ to
construct globally defined invariant solutions of equation \eqref{gHS_alpha}.

To simplify notation, we put $\beta = (\alpha+2)^{-1}$, $\alpha \neq -2$, so equation \eqref{gHS_alpha} gets the form
\begin{equation}
u_{tx} = u\,u_{xx}+\frac{1}{\alpha+2}\,u_x^2.
\label{gHS}
\end{equation}

\section{Preliminaries}

The presentation in this section closely follows
\cite{KrasilshchikVerbovetsky2011}---\cite{KrasilshchikVinogradov1989}, \cite{VK1999}.
Let \,$\pi \colon \mathbb{R}^n \times \mathbb{R}^m \rightarrow \mathbb{R}^n$,
\,$\pi \colon (x^1, \dots, x^n, u^1, \dots, u^m)$
$\mapsto$ $(x^1, \dots, x^n)$,
be a trivial bundle, and
$J^\infty(\pi)$ be the bundle of its jets of the infinite order. The local coordinates on $J^\infty(\pi)$ are
$(x^i,u^\alpha,u^\alpha_I)$, where $I=(i_1, \dots, i_n)$ are multi-indices, and for every local section
$f \colon \mathbb{R}^n \rightarrow \mathbb{R}^n \times \mathbb{R}^m$ of $\pi$ the corresponding infinite jet
$j_\infty(f)$ is a section $j_\infty(f) \colon \mathbb{R}^n \rightarrow J^\infty(\pi)$ such that
$u^\alpha_I(j_\infty(f))
=\displaystyle{\frac{\partial ^{\#I} f^\alpha}{\partial x^I}}
=\displaystyle{\frac{\partial ^{i_1+\dots+i_n} f^\alpha}{(\partial x^1)^{i_1}\dots (\partial x^n)^{i_n}}}$.
We put $u^\alpha = u^\alpha_{(0,\dots,0)}$. Also, we will simplify notation in the following way, e.g., in the
case of $n=2$, $m=1$: we denote $x^1 = t$, $x^2= x$
and $u^1_{(i,j)}=u_{{t \dots t}{x \dots x}}$ with $i$  times $t$ and $j$  times $x$,
or $u_{kx}$, $k \in \mathbb{N}$, for $u_{xx..x}$ with $k$ times $x$.

The  vector fields
\[
D_{x^k} = \frac{\partial}{\partial x^k} + \sum \limits_{\# I \ge 0} \sum \limits_{\alpha = 1}^m
u^\alpha_{I+1_{k}}\,\frac{\partial}{\partial u^\alpha_I},
\qquad k \in \{1,\dots,n\},
\]
$(i_1,\dots, i_k,\dots, i_n)+1_k = (i_1,\dots, i_k+1,\dots, i_n)$,  are called {\it total derivatives}.
They com\-mu\-te everywhere on
$J^\infty(\pi)$.

The {\it evolutionary vector field} associated to an arbitrary vector-valued smooth function
$\varphi \colon J^\infty(\pi) \rightarrow \mathbb{R}^m $ is the vector field
\begin{equation}
\mathbf{E}_{\varphi} = \sum \limits_{\# I \ge 0} \sum \limits_{\alpha = 1}^m
D_I(\varphi^\alpha)\,\frac{\partial}{\partial u^\alpha_I}
\label{evolution_derivation}
\end{equation}
with $D_I=D_{(i_1,\dots\,i_n)} =D^{i_1}_{x^1} \circ \dots \circ D^{i_n}_{x^n}$.
Notice that
\begin{equation}
[\mathbf{E}_{\varphi}, D_{x^i}]=0
\label{commutator_E_phi_D_x}
\end{equation}
for any $\phi$ and $i$.

A system of {\sc pde}s $F_r(x^i,u^\alpha_I) = 0$ of the order $s \ge 1$ with $\# I \le s$,
$r \in \{1,\dots, R\}$ for some $R \ge 1$,
defines the submanifold
$\EuScript{E}=\{(x^i,u^\alpha_I)\in J^\infty(\pi)\,\,\vert\,\,D_K(F_r(x^i,u^\alpha_I))=0,\,\,\# K\ge 0\}$
in $J^\infty(\pi)$.

A function $\varphi \colon J^\infty(\pi) \rightarrow \mathbb{R}^m$ is called a {\it (generator of an
infinitesimal) symmetry} of equation $\EuScript{E}$ when $\mathbf{E}_{\varphi}(F) = 0$ on $\EuScript{E}$. The
symmetry $\varphi$ is a solution to the {\it defining system}
\begin{equation}
\ell_{\EuScript{E}}(\varphi) = 0,
\label{defining_eqns}
\end{equation}
where $\ell_{\EuScript{E}} = \ell_F \vert_{\EuScript{E}}$ with the matrix differential operator
\[
\ell_F = \left(\sum \limits_{\# I \ge 0}\frac{\partial F_r}{\partial u^\alpha_I}\,D_I\right).
\]
The {\it symmetry algebra} $\mathrm{Sym} (\EuScript{E})$ of equation $\EuScript{E}$ is the linear space of
solutions to  (\ref{defining_eqns}) endowed with the structure of a Lie algebra over $\mathbb{R}$ by the
{\it Jacobi bracket} $\{\varphi,\psi\} = \mathbf{E}_{\varphi}(\psi) - \mathbf{E}_{\psi}(\varphi)$.
The {\it algebra of contact symmetries} $\mathrm{Sym}_0 (\EuScript{E})$ is the Lie subalgebra of
$\mathrm{Sym} (\EuScript{E})$ defined as $\mathrm{Sym} (\EuScript{E}) \cap C^\infty(J^1(\pi))$.

Let the linear space $\EuScript{W}$ be either $\mathbb{R}^N$ for some $N \ge 1$ or  $\mathbb{R}^\infty$
endowed  with  local coordinates $w^s$, $s \in \{1, \dots , N\}$ or  $s \in  \mathbb{N}$, respectively.
Locally, a {\it differential covering} of $\EuScript{E}$ is a trivial bundle
$\tau \colon J^\infty(\pi) \times \EuScript{W} \rightarrow J^\infty(\pi)$
equipped with {\it extended total derivatives}
\[
\widetilde{D}_{x^k} = D_{x^k} + \sum \limits_{s}
T^s_k(x^i,u^\alpha_I,w^j)\,\frac{\partial }{\partial w^s}
\]
such that $[\widetilde{D}_{x^i}, \widetilde{D}_{x^j}]=0$ for all $i \not = j$ if and only if
$(x^i,u^\alpha_I) \in \EuScript{E}$. Define
the partial derivatives of $w^s$ by  $w^s_{x^k} =  \widetilde{D}_{x^k}(w^s)$.  This yields the system
\begin{equation}
w^s_{x^k} = T^s_k(x^i,u^\alpha_I,w^j)
\label{WE_prolongation_eqns}
\end{equation}
that is compatible iff $(x^i,u^\alpha_I) \in \EuScript{E}$.
System \eqref{WE_prolongation_eqns} is referred to as the {\it covering equations}
or the {\it Lax representation} of equation $\EuScript{E}$.

\vskip 7 pt

\addtocounter{example_counter}{1}

\noindent
{\sc  Example \arabic{example_counter}.}
A differential covering  for the Hunter--Saxton equation \eqref{HS_eq} has been presented in \cite{Reyes2002}.
In a slightly different notation this is defined on $J^{\infty}(\pi) \times \mathbb{R}$
with $\pi \colon \mathbb{R}^2\times \mathbb{R} \rightarrow \mathbb{R}^2$,
$\pi \colon (t,x,u) \mapsto (t,x)$, by the vector fields
\begin{equation}
\left\{
\begin{array}{lcl}
\tilde{D}_t &=&
\displaystyle{
D_t + \left(\left(\lambda\,u +\frac{1}{2} \right)\,w^2+u_x\,w-u\,u_{xx}\right)\,\frac{\partial}{\partial w},
}
\\
\tilde{D}_x &=&
\displaystyle{
D_x + \left(\lambda\,w^2 - u_{xx}\right)\,\frac{\partial}{\partial w},
}
\end{array}
\right.
\label{HS_total_derivatives}
\end{equation}
or by the system of the covering equations
\begin{equation}
\left\{
\begin{array}{lcl}
w_t &=& \displaystyle{
\left(\lambda\,u +\frac{1}{2} \right)\,w^2+u_x\,w-u\,u_{xx}},
\\
w_x &=& \displaystyle{\lambda\,w^2 - u_{xx}}.
\end{array}
\right.
\label{HS_covering}
\end{equation}
The compatibility condition $(w_t)_x=(w_x)_t$ of this system coincides with equation \eqref{HS_eq}.
The $\tau$-vertical parts of the right-hand sides of \eqref{HS_total_derivatives} are
linear combinations of the vector fields $\partial_w$, $w\,\partial_w$, and $w^2\,\partial_w$.
These vector fields generate the Lie algebra $\mathfrak{sl}_2(\mathbb{R})$ referred to as the
{\it universal algebra} of the covering, \cite{KrasilshchikVinogradov1984}.
\hfill $\diamond$

\vskip 7 pt

Consider operator $\widetilde{\mathbf{E}}_\phi$ obtained by replacing $D_{x^k}$ to
$\widetilde{D}_{x^k}$ in \eqref{evolution_derivation}.
Solutions $\phi=\phi(x^i,u^\alpha_I,w^j)$ to equation $\tilde{\mathbf{E}}_\phi(F) =0$
are referred to as {\it shadows of symmetries in the covering} $\tau$, or just {\it shadows}.

A {\sc pde} $\EuScript{E}$  has two important coverings: the tangent covering $\EuScript{TE}$ and the cotangent
covering $\EuScript{T^{*}E}$. Their covering equations are given by systems $\ell_{\EuScript{E}}(q) = 0$ and
$\ell_{\EuScript{E}}^{*}(p) = 0$, respectively, where $\ell_{\EuScript{E}}^{*}$ is the adjoint operator to
$\ell_{\EuScript{E}}$. The local sections of the tangent covering are (generators of) symmetries, while the
local sections of the cotangent covering are referred to as {\it cosymmetries}. The cosymmetries generate
conservation laws for equation $\EuScript{E}$, see discussion in \cite[Ch.~1]{KrasilshchikVerbovetskyVitolo2017}.
and Example 4 below.

\vskip 7 pt

\addtocounter{example_counter}{1}

\noindent
{\sc  Example \arabic{example_counter}.}
The covering equations for the tangent and cotangent coverings of equation \eqref{gHS} have the form
\begin{equation}
\ell_{\EuScript{E}}(q) =q_{tx} - u\,q_{xx}-\frac{2}{\alpha+2}\,u_x\,q_x- u_{xx}\,q =0.
\label{tangent_covering}
\end{equation}
and
\begin{equation}
\ell_{\EuScript{E}}^{*}(p) =  p_{tx} - u\,p_{xx}-\frac{2\,(\alpha+1)}{\alpha+2}\,(u_x\,p_x- u_{xx}\,p) =0.
\label{cotangent_covering}
\end{equation}
For $\alpha =0$ equations
\eqref{tangent_covering} and \eqref{cotangent_covering} coincide. This property holds for each {\sc pde} that
admits a Lagrangian formulation, \cite[Example~10.1]{KrasilshchikVerbovetskyVitolo2017}.
\hfill $\diamond$
\vskip 7 pt

A {\it recursion operator for symmetries} of a {\sc pde} $\EuScript{E}$ is a B{\"a}cklund autotransformation in
the tangent covering  $\EuScript{TE}$. In other words, this is an operator $\EuScript{R}$ such that
\begin{equation}
\ell_{\EuScript{E}} \circ \EuScript{R} = \EuScript{S} \circ \ell_{\EuScript{E}}
\label{def_recursion_operator_for_symmetries}
\end{equation}
for some operator $\EuScript{S}$. Likewise, a {\it recursion operator for co\-sym\-met\-ri\-es} of a {\sc pde}
$\EuScript{E}$ is a B{\"a}cklund autotransformation in the cotangent covering  $\EuScript{T^{*}E}$.
Taking adjoint operators to both sides of \eqref{def_recursion_operator_for_symmetries} we get
\begin{equation}
\EuScript{R}^{*} \circ \ell^{*}_{\EuScript{E}}  = \ell_{\EuScript{E}}^{*} \circ \EuScript{S}^{*}.
\label{def_recursion_operator_for_cosymmetries}
\end{equation}
Therefore operator $\EuScript{S}^{*}$ is a recursion operator for cosymmetries.

\section{Lax representation}
\label{Lax_representation_section}

Based on Example 1 we conjecture that the generalized Hun\-ter--Saxton equation \eqref{gHS} admits a Lax
representation with the same universal algebra $\mathfrak{sl}_2(\mathbb{R})$. We assume also that the
coefficients of the covering equations are functions of $u$, $u_x$, and $u_{xx}$, that is, there exists
system
\begin{equation}
\left\{
\begin{array}{lcl}
w_t &=& \displaystyle{
T_2\,w^2+T_1\,w+T_0},
\\
w_x &=& \displaystyle{X_2\,w^2 +X_1\,w+X_0},
\end{array}
\right.
\label{gHS_hint_covering}
\end{equation}
with $T_i=T_i(u,u_x,u_{xx})$ and $X_i=X_i(u,u_x,u_{xx})$ such that \eqref{gHS} coincides with the integrability
conditions of \eqref{gHS_hint_covering}. Direct computations give such a system:
\begin{equation}
\left\{
\begin{array}{lcl}
w_t &=& \displaystyle{
\left(\lambda\,u \,u_x^{\alpha} +\frac{1}{\alpha+2} \right)\,w^2+\frac{2}{\alpha+2}\,u_x\,w-u\,u_{xx}},
\\
w_x &=& \displaystyle{\lambda\,u_x^\alpha\,w^2 - u_{xx}}.
\end{array}
\right.
\label{gHS_covering}
\end{equation}
When $\alpha=0$, this system  coincides with  \eqref{HS_covering}. The parameter $\lambda \neq 0$ in both
systems \eqref{HS_covering} and \eqref{gHS_covering}  is non-removable. In accordance with
\cite[\S\S 3.2, 3.6]{KrasilshchikVinogradov1989}, \cite{Krasilshchik2000,IgoninKrasilshchik2000},
to prove this assertion it is sufficient to note that symmetry $V = x\,\partial_x+u\,\partial_u$ of equation
\eqref{gHS} does not admit a lift to a symmetry of system \eqref{gHS_covering}. Therefore the action
$\mathrm{e}^{\epsilon V} \colon (t, x, u, u_t, u_x, u_{xx}, w, w_t, w_x) \mapsto$
$(t, \mathrm{e}^\epsilon x, \mathrm{e}^\epsilon u, \mathrm{e}^\epsilon u_t, u_x, \mathrm{e}^{-\epsilon} u_{xx},
w, w_t, \mathrm{e}^{-\epsilon} w_x)$
of operator $\mathrm{e}^{\epsilon V}$ transforms system \eqref{gHS_covering} with $\lambda = 1$ to system
\[
\left\{
\begin{array}{lcl}
w_t &=& \displaystyle{
\left(\mathrm{e}^\epsilon\,u \,u_x^{\alpha} +\frac{1}{\alpha+2} \right)\,w^2+\frac{2}{\alpha+2}\,u_x\,w-u\,u_{xx}},
\\
\mathrm{e}^{-\epsilon} w_x &=& \displaystyle{ u_x^\alpha\,w^2 - \mathrm{e}^{-\epsilon} u_{xx}},
\end{array}
\right.
\]
which coincides with \eqref{gHS_covering} when $\lambda = \mathrm{e}^\epsilon$.

The map
\[
\partial_w \mapsto -\left(
\begin{array}{lr}
0 & 1
\\
0 & 0
\end{array}
\right),
\,\,
w\,\partial_w \mapsto \frac{1}{2}\,\left(
\begin{array}{lr}
1 & 0
\\
0 & -1
\end{array}
\right),
\,\,
w^2\,\partial_w \mapsto \frac{1}{2}\,\left(
\begin{array}{lr}
0 & 0
\\
1 & 0
\end{array}
\right)
\]
rearranges the Lax representation  \eqref{gHS_covering} into the matrix
form\footnote[1]{this form of a Lax representation is called {\it the zero curvature representation}}
$A_t-B_x=[A,B]$ with
\[
A =
\left(
\begin{array}{lcl}
\displaystyle{0}
&&
\displaystyle{u_{xx}}
\\
\displaystyle{\lambda\,u_x^\alpha}
&&
\displaystyle{0}
\end{array}
\right),
\quad
B =
\left(
\begin{array}{lcl}
\displaystyle{\frac{1}{\alpha+2} \,u_x}
&&
\displaystyle{u\,u_{xx}}
\\
\displaystyle{\lambda\,u\,u_x^\alpha+ \frac{1}{\alpha+2}}
&&
\displaystyle{-\frac{1}{\alpha+2}\,u_x}
\end{array}
\right).
\]

System \eqref{gHS_covering} can be written in the form of the pseudospherical type surface equations
\[
\left\{
\begin{array}{lcl}
d\omega_1 &=& \omega_3 \wedge \omega_2,
\\
d\omega_2 &=& \omega_1 \wedge \omega_3,
\\
d\omega_3 &=& \omega_1 \wedge \omega_2,
\end{array}
\right.
\]
with
\[
\omega_1 = \frac{2}{\alpha+2}\,u_x\,dx,
\]
\[
\omega_2 =
\left(u\,u_{xx}+ \frac{1}{\alpha+2} + \lambda\,u\,u_x^\alpha\right) \,dt
+\left(u_{xx}+\lambda\,u_x^\alpha\right)\,dx,
\]
\[
\omega_3 =
\left(u\,u_{xx}+ \frac{1}{\alpha+2} - \lambda\,u\,u_x^\alpha\right) \,dt
+\left(u_{xx}-\lambda\,u_x^\alpha\right)\,dx,
\]
see discussion of the pseudospherical type equations in
\cite{Sasaki1979,ChernTeneblat1986,Reyes2002,GorkaReyes2012} and references therein.

\section{Contact symmetries}
\label{contact_symmetries_section}

The Lie algebra $\mathrm{Sym}_0(\EuScript{E})$ of contact symmetries  of equation \eqref{gHS}
is generated by functions\footnote[2]{We carried out computations of generators of contact symmetries,
their
commutator tables, shadows of symmetries, and cosymmetries  in the Jets software \cite{Jets}.}
\begin{equation}
\begin{array}{lcl}
\phi_{0,0} &=& x\,u_x-u,
\\
\phi_{1,0} &=& u_t,
\\
\phi_{1,1} &=& -2\,t\,u_t -(\alpha+2)\,x\,u_x+\alpha\,u,
\\
\phi_{1,2} &=& - t^2 \,u_t -(\alpha +2)\,t\,x\,u_x + \alpha\,t\,u -(\alpha+2)\,x,
\end{array}
\label{phis}
\end{equation}
and the family of solutions $U=U(t,u_x)$ to the linear {\sc pde}
\begin{equation}
U_{tu_x} = - \frac{1}{\alpha+2} \,u_x^2 \,U_{u_xu_x} - u_{x}\,U_{u_x} +U.
\label{U_defining_eq}
\end{equation}
The commutator table of $\mathrm{Sym}_0(\EuScript{E})$ is given by equations
\[
\fl
\begin{array}{lcl}
\{\phi_{0,0}, \phi_{1,i} \} &=& 0,
\\
\{\phi_{0,0}, U\} &=& U,
\\
\{\phi_{1,0}, \phi_{1,1}\} &=& 2\,\phi_{1,0},
\\
\{\phi_{1,0}, \phi_{1,2}\} &=& -\phi_{1,1},
\\
\{\phi_{1,0}, U\} &=&- U_t,
\\
\{\phi_{1,1}, \phi_{1,2}\} &=& 2\,\phi_{1,2},
\\
\{\phi_{1,1}, U\} &=& 2\,t\, U_t-2\,u_x\,U_{u_x} - \alpha U,
\\
\{\phi_{1,2}, U\} &=& t^2\, U_t - \alpha t\,U-(2\,t\,u_x+\alpha+2)\,U_{u_x},
\\
\{U, \tilde{U}\}&=& 0.
\end{array}
\]
This table implies that the contact symmetry algebra of \eqref{gHS} is the semi-direct sum
$\mathrm{Sym}_0(\EuScript{E}) = \mathfrak{s}_4 \ltimes \mathfrak{a}_\infty$
of the four-di\-men\-si\-o\-nal subalgebra
$\mathfrak{s}_4 = \langle \, \phi_{0,0},  \phi_{1,0}, \phi_{1,1}, \phi_{1,0} \,\rangle$
that is isomorphic to $\mathfrak{gl}_2(\mathbb{R})$,
and the infinite-dimensional Abelian ideal
$\mathfrak{a}_\infty$ spanned by solutions to \eqref{U_defining_eq}.

Equation \eqref{U_defining_eq} has solutions of the form $\psi(A) = A\,u_x+A^{\prime}$. These
functions generate a sub-ideal $\mathfrak{b}_\infty \subsetneq  \mathfrak{a}_\infty$.
The action of $\mathfrak{s}_4$ on $\mathfrak{b}_\infty$ is given by equations
\[
\fl
\begin{array}{lcl}
\{\phi_{0,0}, \psi(A)\} &=& \psi(A),
\\
\{\phi_{1,0}, \psi(A)\} &=& \psi(-A_t),
\\
\{\phi_{1,1}, \psi(A)\} &=& \psi(2\,t\,A_t-(\alpha+2) \,A),
\\
\{\phi_{1,2}, \psi(A)\} &=& \psi(t^2\,A_t-(\alpha+2)\,t \,A).
\end{array}
\]
This action has the following reformulation: the vector space $\mathbb{A}$ of smooth functions
$A=A(t)$ has a $\mathfrak{s}_4$--module structure
\begin{equation}
\rho \colon \mathfrak{s}_4 \times \mathbb{A} \rightarrow \mathbb{A},
\qquad
\rho \colon (\phi, A) \mapsto \phi \centerdot A
\label{rho_def}
\end{equation}
defined by formulae
\begin{equation}
\fl
\begin{array}{lclclcl}
\phi_{0,0} \centerdot A &=& A,
&\hspace{10pt}&
\phi_{1,1} \centerdot A &=& 2\,t\,A_t-(\alpha+2) \,A,
\\
\phi_{1,0} \centerdot A &=& -A_t,
&&
\phi_{1,2} \centerdot A &=& t^2\,A_t-(\alpha+2)\,t \,A.
\end{array}
\label{rho_def_formulae}
\end{equation}

\vskip 7 pt
\addtocounter{remark_counter}{1}

\noindent
{\sc  Remark \arabic{remark_counter}.}
The problem to find all the local symmetries of the form $U(t,u_x)$ is
as hard as the problem to  find all solutions to equation \eqref{gHS}, since equation \eqref{U_defining_eq} is
contact--equivalent to \eqref{gHS}. The proof of this statement mimics the proof of contact equivalence of
equations \eqref{gHS} and \eqref{EP} presented in \cite{Morozov2005}.
\hfill $\diamond$

\section{Recursion operators}
\label{recursion_operators_section}

In this section we use the methods of
\cite{KrasilshchikKersten1994,KrasilshchikKersten1995,KrasilshchikKersten2000,KrasilshchikVerbovetskyVitolo2017}
to find local and nonlocal recursion operators for symmetries of equation \eqref{gHS}.

To construct local recursion operators of first order
we search for shadows of symmetries of the form
\[
\sigma = Q_1\,q_t+Q_2\,q_x+Q_3\,q,
\quad
Q_i=Q_i(t,x,u,u_t,u_x,u_{tt},u_{xx}),
\]
where $q$ is a solution to  \eqref{tangent_covering}.
Direct computations then give the following shadows:
\[
\sigma_0 = -q_t+ \frac{E}{u_{xx}}\,q_x,
\]
\[
\sigma_1 = 2\,t\,q_t - 2\,\frac{t\,E +u_x}{u_{xx}}\,q_x -\alpha\,q,
\]
\[
\sigma_2 =  t^2\,q_t - \frac{t^2\,E+2\,t\,u_x+\alpha+2}{u_{xx}}\,q_x
-\alpha\,t\,q,
\]
where $E$ is the right hand side of equation \eqref{gHS}. Therefore we have
\vskip 7 pt
\noindent
{\sc Proposition 1}. {\it Differential operators
\begin{equation}
\fl
\EuScript{R}_0 =-D_t+ \frac{E}{u_{xx}}\,D_x,
\label{R_0}
\end{equation}
\begin{equation}
\fl
\EuScript{R}_1 =2\,t\,D_t - 2\,\frac{t\,E +u_x}{u_{xx}}\,D_x -\alpha,
\label{R_1}
\end{equation}
\begin{equation}
\fl
\EuScript{R}_2  = t^2 D_t
- \frac{t^2\,E+2\,t\,u_x+\alpha+2}{u_{xx}}D_x
-\alpha t
\label{R_2}
\end{equation}
provide local recursion operators for symmetries of $\EuScript{E}$.
}
\vskip 5 pt
\noindent
{\bf Proof} follows from  the general results of \cite{KrasilshchikVerbovetskyVitolo2017}, or from
identities
\[
[\ell_{\EuScript{E}},\EuScript{R}_0] =
-\frac{u_x\,(u_x\,u_{xxx}-(\alpha+4)\,u_{xx}^2)}{(\alpha+2)\,u_{xx}}\,\ell_{\EuScript{E}},
\]
\[
[\ell_{\EuScript{E}},\EuScript{R}_1] =
\frac{2\,u_x\,((t\,u_x+\alpha+2)\,u_{xxx}-(\alpha+4)\,t\,u_{xx}^2)}{(\alpha+2)\,u_{xx}^2}\,\ell_{\EuScript{E}},
\]
\[
[\ell_{\EuScript{E}},\EuScript{R}_2] =
\frac{(t\,u_x+\alpha+2)^2\,u_{xxx}-(\alpha+4)\,t^2\,u_x\,u_{xx}^2}{(\alpha+2)\,u_{xx}^2}\,\ell_{\EuScript{E}}.
\]
\hfill $\Box$

\vskip 7 pt
Notice that the local recursion operators have the following commutator table:
\begin{equation}
[\EuScript{R}_0, \EuScript{R}_1] = 2\,\EuScript{R}_0,
\quad
[\EuScript{R}_0, \EuScript{R}_2] = -\EuScript{R}_1,
\quad
[\EuScript{R}_1, \EuScript{R}_2] = 2\,\EuScript{R}_2,
\label{R_commutators}
\end{equation}
in other words, they constitute the Lie algebra $\mathfrak{sl}_2(\mathbb{R})$.

\vskip 7 pt

To find nonlocal recursion operators for symmetries we consider the Whitney product of the tangent covering
\eqref{tangent_covering} and the cotangent covering given by equation \eqref{cotangent_covering}.
Then Green's formula
\[
\left(q\,\ell^{*}_{\EuScript{E}}(p) - p \,\ell_{\EuScript{E}}(q)\right)\,dt\wedge dx
=
d\left(q\,p_x\right)\wedge dx
\]
\[
\qquad
+ dt \wedge
d\left(p\,q_t-u\,p\,q_x- \left(\frac{\alpha}{\alpha+2}\,p\,u_x-u\,p_x\right)\,q \right)
\]
provides the canonical conservation law \cite[p.~22]{KrasilshchikVerbovetskyVitolo2017}
\begin{equation}
\left\{
\begin{array}{lcl}
S_t &=& \displaystyle{p\,q_t-u\,p\,q_x- \left(\frac{\alpha}{\alpha+2}\,p\,u_x-u\,p_x\right)\,q},
\\
S_x &=& q\,p_x.
\end{array}
\right.
\label{W_nonlocality}
\end{equation}
In particular, substituting for the solution $p = u_x^{-2}$ of \eqref{cotangent_covering} into
\eqref{W_nonlocality}
defines nonlocality $s_1$ by equations
\begin{equation}
\left\{
\begin{array}{lcl}
s_{1,t} &=& \displaystyle{\frac{q_t-u\,q_x}{u_x^2}
- \frac{2\,(\alpha+2)\,u\,u_{xx}-\alpha\,u_x^2}{(\alpha+2)\,u_x^3}\,q},
\phantom{\frac{\frac{A}{A}}{\frac{A}{A}}}
\\
s_{1,x} &=& \displaystyle{-2\,\frac{u_{xx}}{u_x^2}\,q}.
\phantom{\frac{\frac{A}{A}}{\frac{A}{A}}}
\end{array}
\right.
\label{w_1_nonlocality}
\end{equation}
Likewise, the cosymmetry $p = u_x^{\alpha+1}$ defines the nonlocality $s_2$ by system
\begin{equation}
\left\{
\begin{array}{lcl}
s_{2,t} &=& \displaystyle{
u_x^{\alpha} \left(u_x(q_t-u\,q_x)
+ (\alpha \,(\alpha+2)^{-1}  u_x^2+(\alpha+1) u u_{xx})\,q\right)
},
\\
s_{2,x} &=& (\alpha+1)\, u_x^\alpha\,u_{xx}\,q.
\end{array}
\right.
\label{w_2_nonlocality}
\end{equation}
We obtain four shadows of symmetries in the tangent covering of the form
$\sigma = Q_1\,q_t+Q_2\,q_x+Q_3\,q+Q_4\,s_1+Q_5\,s_2$ with nontrivial functions $Q_4$ and $Q_5$:
\[
\fl
\sigma_4 =u_x\,s_1 - \frac{t\,u_x+\alpha+2}{(\alpha+2)\,u_x}\,q,
\label{nonlocal_rho_1_1}
\]
\[
\fl
\sigma_5 =(t\,u_x+1)\,s_1
+ \frac{1}{(\alpha+2)^2} t^3\,q_t
-\frac{t^3\,u_x\,E
+3\,t\,u_x\,(t\,u_x+\alpha+2)+(\alpha+2)^2
 }{(\alpha+2)^2\,u_x\,u_{xx}}\,q_x
\]
\[
\fl\quad\quad
-\frac{(2\,\alpha+1)\,t^2\,u_x^2+(\alpha+2)^2\,(t\,u_x+1)}{(\alpha+2)^2\,u_x^2}\,q,
\]
\[
\fl
\sigma_6 = u_x^{-\alpha-2}\,s_2 - \frac{t\,u_x+\alpha+3}{(\alpha+2)\,u_x}\,q,
\]
\[
\fl
\sigma_7 =
\frac{(\alpha+4)\,t\,u_x+(\alpha+2)^2}{(\alpha+2)^2\,u_x^{\alpha+3}}\, s_2 + \frac{t\,u_x+1}{\alpha+2}\,s_1
- \frac{(\alpha+3)\,(t\,u_x+\alpha+2)^2}{(\alpha+2)^3\,u_x^2}\,q.
\]
From the second equation in \eqref{w_1_nonlocality} we have
\[
s_1 = -2\,D_x^{-1}\left(\frac{u_{xx}}{u_x^2}\,q\right),
\]
therefore the nonlocal recursion operators
\[
-2\,u_x\,D_x^{-1}\circ \frac{u_{xx}}{u_x^2} - \frac{t\,u_x+\alpha+2}{(\alpha+2)\,u_x}
\]
and
\[
\fl
-2\,(t\,u_x+1)\,D_x^{-1}\circ \frac{u_{xx}}{u_x^2} + \frac{t^3}{(\alpha+2)^2}\,D_t
-\frac{(2\,\alpha+1)\,t^2\,u_x^2+(\alpha+2)^2\,(t\,u_x+1)}{(\alpha+2)^2u_x^2}\,
\]
\[
\fl\quad\quad
-\frac{t^3\,u_x\,((\alpha+2)\,u\,u_{xx}+u_x^2)+3\,(\alpha+2)\,t\,u_x\,(t\,u_x+\alpha+2)+(\alpha+2)^3
 }{(\alpha+2)^3\,u_x\,u_{xx}}\,D_x
\]
are associated with shadows $\sigma_4$ and $\sigma_5$.
In the same way from the second equation in \eqref{w_2_nonlocality} we have
\[
s_{2} = (\alpha+1)\, D_x^{-1} \left( u_x^\alpha\,u_{xx}\,q\right),
\]
therefore shadows $\sigma_6$ and $\sigma_7$ produce the nonlocal recursion operators
\[
(\alpha+1)\, u_x^{-\alpha-2}\,D_x^{-1}\circ u_x^\alpha\,u_{xx}  - \frac{t\,u_x+\alpha+3}{(\alpha+2)\,u_x}
\]
and
\[
\fl
(\alpha+1)\,\frac{(\alpha+4)\,t\,u_x+(\alpha+2)^2}{(\alpha+2)^2\,u_x^{\alpha+3}}\, D_x^{-1}\circ u_x^\alpha\,u_{xx}
-2\, \frac{t\,u_x+1}{\alpha+2}\,D_x^{-1}\circ \frac{u_{xx}}{u_x^2}
- \frac{(\alpha+3)\,(t\,u_x+\alpha+2)^2}{(\alpha+2)^3\,u_x^2}
\]
respectively.

\section{Higher symmetries}
\label{higher_symmetries_section}

The action of the local recursion operators \eqref{R_0} --- \eqref{R_2} on the contact symmetries
\eqref{phis} produces the Lie subalgebra $\mathfrak{s}_\infty \subset \mathrm{Sym}(\EuScript{E})$. In this
section we study the structure of $\mathfrak{s}_\infty$. We have
$\phi_{1,i} = \EuScript{R}_i(\phi_{0,0})$ for $i \in \{0, 1, 2\}$,
hence
\begin{equation}
\mathfrak{s}_\infty = \langle \, \EuScript{R}^{(p,q,r)}(\phi_{0,0})
\,\,
\vert
\,\, p, q, r \in \mathbb{N} \cup \{0\}\,\rangle
\label{g_infty_def}
\end{equation}
for
\[
\EuScript{R}^{(p,q,r)}=\EuScript{R}^p_0 \circ \EuScript{R}^q_1\circ \EuScript{R}^r_2
=
\underbrace{\EuScript{R}_0 \circ \dots \circ \EuScript{R}_0}_{p~ \mathrm{times}}
\circ
\underbrace{\EuScript{R}_1 \circ \dots \circ \EuScript{R}_1}_{q~ \mathrm{times}}
\circ
\underbrace{\EuScript{R}_2 \circ \dots \circ \EuScript{R}_2}_{r~ \mathrm{times}}
\]
due to  \eqref{R_commutators}.

\vskip 7 pt
\noindent
{\sc Lemma}.
{\it
For every $\phi \in C^{\infty}(\EuScript{E})$ and $i \in \{0, 1, 2\}$ there holds}
\[
\{\EuScript{R}_i(\phi), \psi(A) \}  =  \EuScript{R}_i \left( \{\phi, \psi(A) \} \right).
\]
Proof. Suppose $i= 0$. Denote $W = \EuScript{R}_0(x)$, so $\EuScript{R}_0 = -D_t+W\,D_x$,
and for arbitrary $A=A(t)$ denote $\psi=\psi(A)$ for short. Using \eqref{commutator_E_phi_D_x} we have
\[
\{\EuScript{R}_0(\phi), \psi \}  -  \EuScript{R}_0 \left( \{\phi, \psi \} \right) =
\mathbf{E}_{-D_t(\phi)+W\,D_x(\phi)}(\psi) - \mathbf{E}_{\psi}(-D_t(\phi)+W\,D_x(\phi))
\]
\[
\qquad\qquad
+ D_t(\mathbf{E}_{\phi}(\psi)-\mathbf{E}_{\psi}(\phi)) - W\,D_x(\mathbf{E}_{\phi}(\psi)-\mathbf{E}_{\psi}(\phi))
\]
\[
\fl
\quad\qquad
=A \, D_x(-D_t(\phi) + W\,D_x(\phi))-\mathbf{E}_{\psi}(w)\,D_x(\phi) +[\mathbf{E}_\psi,D_t](\phi)
-W\,[\mathbf{E}_\psi,D_x](\phi)
\]
\[
\qquad\qquad
+D_t(A\,D_x(\phi))-W\,A\,D_x^2(\phi) =
\left(A\,D_x(W)+A^{\prime} - \mathbf{E}_\psi(W)\right) \,D_x(\phi) =0,
\]
since direct computations give  $\mathbf{E}_\psi(W) = A\,D_x(W)+A^{\prime}$.

For $i=1$ and $i=2$ the proof is similar.
\hfill $\Box$
\vskip 7 pt

In particular, we have (recall notation of \eqref{rho_def}, \eqref{rho_def_formulae})
\[
\EuScript{R}_i(\psi(A)) = \EuScript{R}_i (\{\phi_{0,0}, \psi(A)\}) =
\{\phi_{1,i}, \psi(A)\} = \psi(\phi_{1,i} \centerdot A).
\]
Combining this with \eqref{g_infty_def} we obtain
\vskip 7 pt

\noindent
{\sc Proposition 2}.
{\it
Representation \eqref{rho_def}, \eqref{rho_def_formulae} admits a prolongation to the Lie algebra
$\mathfrak{s}_\infty$  gi\-ven by formula}
\begin{equation}
\fl
\EuScript{R}^{(p,q,r)}(\phi_{0,0}) \centerdot A
=
\underbrace{\phi_{1,0} \centerdot  (\dots (\phi_{1,0}\, \centerdot \,\,}_{p~ \mathrm{times}}
\underbrace{(\phi_{1,1} \centerdot  (\dots (\phi_{1,1}\, \centerdot \,\,}_{q~ \mathrm{times}}
\underbrace{(\phi_{1,2} \centerdot  (\dots (\phi_{1,2}\, \centerdot}_{r~ \mathrm{times}}
A)\dots).
\label{rho_prolongation}
\end{equation}
\hfill $\Box$

\vskip 7 pt

To show that $\mathfrak{s}_\infty$ has the structure of the so-called {\it Lie algebra of matrices of
the complex size} introduced in \cite{Feigin1988}  and studied in \cite{PostHijligenberg1996},
we recall the constructions of the last paper.

Let $\mathfrak{d}$ denote the Lie algebra of differential operators of the form
$p_n(t)\,\partial^n_t + p_{n-1}(t)\,\partial^{n-1}_t +
\dots + p_1(t)\,\partial_t +p_0(t)$
where $p_k \in \mathbb{C}[t]$ for $k \in \{0, \dots, n\}$ and
$n \in \mathbb{N} \cup \{0\}$, with the Lie bracket defined by the commutator.
For fixed $\lambda \in \mathbb{C}$ consider the subalgebra
$\mathfrak{gl}(\lambda) \subset \mathfrak{d}$ generated by the differential operators
$1$,
\begin{equation}
T_0=-\partial_t,
\,\,
T_1=2\,t\,\partial_t-\lambda+1,
\,\,
T_2= t^2\,\partial_t-(\lambda-1)\,t.
\label{T_def}
\end{equation}
The Lie algebra $\mathfrak{gl}(\lambda)$
is isomorphic to $\EuScript{U}(\mathfrak{sl}_2(\mathbb{C}))/I_{\lambda}$,
where $\EuScript{U}(\mathfrak{sl}_2(\mathbb{C}))$ is the universal enveloping algebra
of $\mathfrak{sl}_2(\mathbb{C})$ and $I_\lambda$ is the ideal in $\EuScript{U}(\mathfrak{sl}_2(\mathbb{C}))$
generated by the differential operator $2\,T_2\circ T_0 +2\,T_0\circ T_2 + T_1\circ T_1 + (\lambda -1)^2$.

Comparing
\eqref{rho_def_formulae}, \eqref{rho_prolongation}, and
\eqref{T_def}, we obtain the following statement:
\vskip 7 pt
\noindent
{\sc Theorem}.
{\it
The Lie algebra
 $\mathfrak{s}_\infty \subset \mathrm{Sym}(\EuScript{E})$ is isomorphic to $\mathfrak{gl}(\alpha+3)$.
}
\hfill $\Box$
\vskip 7 pt

The results of
\cite[\S~4.1]{PostHijligenberg1996} yield

\vskip 7 pt
\noindent
{\sc Corollary.}
{\it
The Lie algebra $\mathfrak{s}_\infty$ has a basis given by symmetries
$\phi_{0,0}$, $\phi_{1,0}$, $\phi_{1,1}$, $\phi_{1,2}$, and
$\phi_{n,2 n-k} = \mathrm{ad}^k_{\phi_{1,0}} \left(\EuScript{R}_2^n(\phi_{0,0})\right)$,
$n \ge 2$, $k \in \{0, \dots, 2\,n\}$.
}
\hfill $\Box$

\vskip 7 pt
\addtocounter{example_counter}{1}

\noindent
{\sc  Example \arabic{example_counter}.}
Symmetries $\phi_{2,0}$, ... , $\phi_{2,4}$ are given by equations
\[\fl
\phi_{2,0}=24\,(u_{tt}-u_{xx}^{-1}\,E^2),
\]
\[\fl
\phi_{2,1}=-24\,\left(
t\,u_{tt}-\frac{1}{2}\,(\alpha-1)\,u_t
-\frac{ E\,(t+u_x)}{u_{xx}}
\right),
\]
\[\fl
\phi_{2,2} = 12\,\left(t^2\,u_{tt}-(\alpha-1)\,t\,u_t
-\frac{E\,(t^2\,E+2\,t\,u_x+\alpha+2)}{u_{xx}}\right)
\]
\[\fl
\qquad
+2\,((\alpha+1)\,(\alpha+2)\,x\,u_x-(\alpha^2+3\,\alpha+8)\,u),
\]
\[\fl
\phi_{2,3} = -4\,t^3\,u_{tt}+6\,(\alpha-1)\,t^2\,u_t
-2\,(\alpha^2+3\,\alpha+8)\,t\,u
+2\,(\alpha+1)\,(\alpha+2)\,x\,(t\,u_x+1)
\]
\[\fl
\qquad
+4\,u_{xx}^{-1}\,\left(
t^3\,E^2-3\,t\,(t\,u_x+\alpha+2)\,E+(\alpha+2)\,u_x
\right),
\]
\[\fl
\phi_{2,4}= t^4\,u_{tt}+2\,(\alpha-1)\,t^3\,u_t
-\frac{t^2\,E\,
(t^2\,E^3+2\,(2\,t\,u_x+3\,(\alpha+2)))+(\alpha+2)\,(4\,t\,u_x+\alpha+2)}{u_{xx}}
\]
\[\fl
\qquad
-(\alpha+1)\,(\alpha+2)\,t\,x\,(t\,u_x+2)
+(\alpha^2+3\,\alpha+8)\,t^2\,u.
\]
\hfill $\diamond$

\vskip 5 pt

\addtocounter{remark_counter}{1}

\noindent
{\sc  Remark \arabic{remark_counter}.}
When $\alpha=0$,
the Lie algebra $\mathfrak{s}_\infty$ is a proper subalgebra of
$\mathrm{Sym}(\EuScript{E})$. The family of symmetries of third order
\begin{equation}
\eta_{m} = \frac{u_{xx}^m}{u_x^{4\,m+7}}\,\left((m+2)\,u_x\,u_{xxx} - (2\,m+3)\,u_{xx}^2\right),
\,\,
m  \in \mathbb{R},
\label{third_order_symmetries}
\end{equation}
was found in \cite{Wang2010}. We have $\eta_{m} \not \in \mathfrak{s}_\infty$.
The action of the local recursion operators $\EuScript{R}_i$ on $\eta_{m}$ provides a family of
higher symmetries  of increasing order. This family is not included in $\mathfrak{s}_\infty$.

We have no examples of higher symmetries that are not included in $\mathfrak{s}_\infty$ when  $\alpha \neq 0$.
\hfill $\diamond$

\vskip 7 pt
\addtocounter{remark_counter}{1}

\noindent
{\sc  Remark \arabic{remark_counter}.}
The local recursion operators $\EuScript{R}_i$ preserve the ideal $\mathfrak{a}_\infty$, since
$\EuScript{R}_i$
map so\-lu\-ti\-ons of equation \eqref{U_defining_eq} to solutions of the same equation,
\hfill $\diamond$

\vskip 7 pt

\addtocounter{remark_counter}{1}

\noindent
{\sc  Remark \arabic{remark_counter}.}
When $\alpha=0$, the family of local recursion operators of fourth order
$\EuScript{P}_m = \EuScript{P}_{m,1} \circ \EuScript{P}_{m,2} \circ \EuScript{P}_{m,3} \circ D_x$
with
\[
\EuScript{P}_{m,1} =
D_x + \frac{u_{xxx}}{u_{xx}},
\quad
\EuScript{P}_{m,2} =
\frac{4\,u_{xx}^2-u_x\,u_{xxx}}{u_x^{4 m -5} \,u_{xx}^{5-m}}\,D_x
+ D_x \circ \frac{4\,u_{xx}^2-u_x\,u_{xxx}}{u_x^{4 m -5} \,u_{xx}^{5-m}},
\quad
\EuScript{P}_{m,3} =
D_x - \frac{u_{xxx}}{u_{xx}}
\]
was  constructed in \cite{Wang2010}. We have $\EuScript{P}_m(\psi(A)) = 0$,  hence $\EuScript{P}_m$ is not a
linear combination of the recursion operators of the form
$\EuScript{R}^{(p,q,r)}$.

We have no examples of local recursion operators
that do not belong to the span of $\EuScript{R}^{(p,q,r)}$ when $\alpha \neq 0$.
\hfill $\diamond$

\section{Cosymmetries}
\label{cosymmetries_section}

Equations \eqref{tangent_covering} and \eqref{cotangent_covering} coincide when $\alpha =0$, hence in this case
cosymmetries are the same as the generators of symmetries. For other values of  $\alpha$ we have
\vskip 5 pt
\noindent
{\sc Proposition 3}.
{\it
All the cosymmetries of equation \eqref{gHS} with $\alpha \neq 0$ that belong to $C^{\infty}(J^1(\pi))$ have the
form $\psi=V(t,u_x)$, where functions $V$ are solutions to the {\sc pde}}
\begin{equation}
V_{tu_x} + \frac{1}{\alpha+2}\, u_x^2\, V_{u_xu_x} + \frac{2-\alpha}{\alpha+2}\,u_x\,V_{u_x}
- \frac{2\,(\alpha+1)}{\alpha+2}\,V =0,
\label{V_cosymmetry_equation}
\end{equation}
which is the adjoint equation for \eqref{U_defining_eq}.
\hfill $\Box$

\vskip 7 pt
Equation \eqref{V_cosymmetry_equation} is equivalent to equation \eqref{gHS} w.r.t. the pseudogroup of contact
transformations on $J^2(\pi)$. The proof of this assertion is similar to the proof of contact equivalence of
equations \eqref{gHS} and \eqref{EP} given in \cite{Morozov2005}. Therefore the problem to find all solutions to
equation \eqref{V_cosymmetry_equation} is as hard as the problem to find all solutions to equation \eqref{gHS}.
Nevertheless, we can find some particular solutions of \eqref{V_cosymmetry_equation}. For example, when $V_t=0$,
this equation get the form of a linear ordinary differential equation of second order. The general solution of
this {\sc ode} is a linear combination with constant coefficients of two fundamental solutions
$\psi_1 = u_x^{-2}$ and $\psi_2 = u_x^{\alpha+1}$.

Equation \eqref{gHS} has higher cosymmetries. E.g., a family of cosymmetries of third order is defined by
formulae
\begin{equation}
\psi_H = u_{xx}^{-\frac{\alpha+6}{\alpha+4}}\,u_{xxx}\,
\left(\int H^{\prime}(z)\,z^{-\frac{\alpha+6}{\alpha+4}} \,dz\right)-(\alpha+4)\,u_x^{\alpha+1}\,H(z),
\qquad z = u_{xx}\,u_x^{-\alpha-4}
\label{third_order_cosymmetries_1}
\end{equation}
when $\alpha \neq -4$ and
\begin{equation}
\psi_G = \frac{
u_x\,G^{\prime}(u_{xx})\,u_{xxx} -2\,u_{xx}\,G(u_{xx})
}{u_x^3\,u_{xx}}
\label{third_order_cosymmetries_2}
\end{equation}
when $\alpha = -4$, where $H$ and $G$ are arbitrary functions of their arguments.

\vskip 7 pt

\addtocounter{remark_counter}{1}

\noindent
{\sc  Remark \arabic{remark_counter}.}
Since symmetries and cosymmetries coincide when $\alpha =0$, equation \eqref{third_order_cosymmetries_1}  with
$\alpha=0$ provides a family of symmetries of third order for equation \eqref{HS_eq}. This family
ge\-ne\-ra\-li\-zes \eqref{third_order_symmetries}. Indeed, $\eta_m$ coincides with $\frac{1}{2}\,\psi_H$ when
$\alpha = 0$ and $H(z) \equiv z^{m+2}$.
\hfill $\diamond$
\vskip 7 pt

Using  \eqref{def_recursion_operator_for_cosymmetries} we find three local recursion operators for cosymmetries
given by equations
\begin{equation}
\EuScript{S}_0=-\EuScript{R}_0,
\qquad
\EuScript{S}_1=-\EuScript{R}_1-2\,\alpha,
\qquad
\EuScript{S}_2=-\EuScript{R}_2-2\,\alpha\,t.
\label{local_recursion_operators_for_cosymmetries}
\end{equation}
We have
$[\EuScript{S}_0, \EuScript{S}_1] = -2\,\EuScript{S}_0$,
$[\EuScript{S}_0, \EuScript{S}_2] = \EuScript{S}_1$,
$[\EuScript{S}_1, \EuScript{S}_2] = -2\,\EuScript{S}_2$,
hence these operators constitute the Lie algebra isomorphic to $\mathfrak{sl}_2(\mathbb{R})$.
The representation of this Lie algebra on the space of solutions to  equation
\eqref{V_cosymmetry_equation} is given by formulae
\[
\EuScript{S}_0(V) = \partial_t\,V,
\qquad
\EuScript{S}_1(V) = -2\,t\,\partial_t\,V+2\,u_x\,\partial_{u_x}\,V-\alpha\,V,
\]
\[
\EuScript{S}_2(V) = -t^2\,\partial_t\,V+(2\,t\,u_x+\alpha+2)\,\partial_{u_x}\,V-\alpha\,t\,V.
\]
Action of operators \eqref{local_recursion_operators_for_cosymmetries} on cosymmetries from families
\eqref{third_order_cosymmetries_1} or \eqref{third_order_cosymmetries_2}  produces cosymmetries of higher order.

\vskip 7 pt

\addtocounter{example_counter}{1}

\noindent
{\sc  Example \arabic{example_counter}.}
Function $\zeta =(u_x\,u_{xxx}-2\,u_{xx}^2)\,u_x^{-3}\,u_{xx}^{-1}$ is a co\-sym\-met\-ry of equation
\eqref{gHS_alpha} for each $\alpha \neq -2$. The related conservation law is
\[
\Omega = \frac{u_x u_{xxx}-u_{xx}^2}{u_x^2 u_{xx}}\,(dx + u\,dt)
- \frac{\ln \vert u_{xx}\vert -(\alpha+4)\,\ln \vert u_x\vert}{\alpha+2}\,dt.
\]
The simple induction shows that
\[
\EuScript{S}_0^k(\zeta)= \frac{u_x^{2(k-1)}}{2^k\,u_{xx}^{k+1}}\,
u_{(k+3)x}+W_{k}(u,u_x,u_{xx}, \dots u_{(k+2)x})
\]
for certain functions $W_k$. Therefore for each $\alpha \neq -2$ equation \eqref{gHS} has cosymmetries of arbitrary
higher order.
\hfill $\diamond$

\section{Invariant solutions}
\label{invariant_solutions_section}

In this section we give an example of implementing higher symmetries for finding globally-defined solutions of
the generalized Hunter--Saxton equation. Namely, we construct $\phi_{2,0}$--invariant solutions to  \eqref{gHS}.
These solutions satisfy the over-determined system that contains \eqref{gHS} and equation
\[
u_{tt} = \frac{(u\,u_{xx}+(\alpha+2)^{-1}\,u_x^2)^2}{u_{xx}}.
\]
The compatibility conditions for this system  get the form
\begin{equation}
u_t = u\,u_x - \frac{u_x^3}{(\alpha+2)^2\,u_{xx}^3}\,(u_{x}\,u_{xxx}-2\,(\alpha+3)\,u_{xx}^2),
\label{compatibility_pde}
\end{equation}
\begin{equation}
u_{xxxx} = \frac{3\,u_{xxx}^2}{u_{xx}}
-\frac{(\alpha+5)\,(2\,u_x\,u_{xx}\,u_{xxx}-(\alpha+4)\,u_{xx}^3}{u_x^2},
\label{compatibility_ode}
\end{equation}
whence to obtain $\phi_{2,0}$--invariant solutions of \eqref{gHS} we proceed as follows: first, we find the
general solution to  {\sc ode} \eqref{compatibility_ode} in the form
$u = S(x, c_1, c_2, c_3, c_4)$. This solution depends
on four arbitrary constants $c_1$, ... , $c_4$. Second, we replace these constants by unknown functions $s_1(t)$, ... ,
$s_4(t)$ and substitute the obtained function $u = S(x,s_1(t),s_2(t),s_3(t),s_4(t))$ into equation
\eqref{compatibility_pde}. This yields a system of {\sc odes} that defines functions $s_i(t)$.

Equation \eqref{compatibility_ode} has four-dimensional solvable algebra of point symmetries generated by vectors
$\partial_x$, $\partial_u$, $x\,\partial_x$, $u\,\partial_u$, therefore this equation is integrable by
quadratures in accordance with the Lie--Bianchi theorem \cite[\S~167]{Bianchi}, \cite[Th.~2.64]{Olver},
\cite{DuzhinLychagin1991}.  Indeed, substituting for $z=u_x$, $z_{x} = w(z)$ yields the {\sc ode} of second
order
\[
w_{zz} = \frac{2\,w_z^2}{w} -\frac{2\,(\alpha+5)\,w_z}{z}+\frac{(\alpha+4)\,(\alpha+5)\,w}{z^2}.
\]
Then we put
$w_z = p(z)\,w$  and reduce the order
of this {\sc ode} by one:
\[
p_z = \left(p-\frac{\alpha+5}{z}\right)^2-\frac{\alpha+5}{z^2}.
\]
New function
$r(z)=p(z)-(\alpha+5)\,z^{-1}$
satisfies the separable {\sc ode}
$r_z=r^2$.
Its general solution $r=(c_0-z)^{-1}$ provides
\[
w_z =\left(\frac{1}{c_0-z}+\frac{\alpha+5}{z}\right)\,w
\]
and therefore $w=c_1\,z^{\alpha+5}\,(c_0-z)^{-1}$,  which gives
\begin{equation}
u_{xx} = c_1\,u_x^{\alpha+5}\,(c_0-u_x)^{-1}.
\label{reduced_ode}
\end{equation}

While this equation is integrable by quadratures, its general solution is too complicated
for arbitrary value of $\alpha$.
We consider case $\alpha =-\frac{7}{2}$ when the formula for the general solution
of equation \eqref{reduced_ode} simplifies enough to give explicit expression for the
globally-defined
$\phi_{2,0}$-- invariant solution
of \eqref{gHS}. We obtain
\[
u =
\frac{s_1}{s_3^2}\,\left((x+s_2)^2+s_3\right)^{\frac{3}{2}}
-\frac{s_1}{s_3^2}\,(x+s_2)^3
-\frac{3\,s_1}{2\,s_3}\,x+s_4,
\]
where $s_1(t)$, ..., $s_4(t)$ are solutions to system
\[
\left\{
\begin{array}{lcl}
s_{1,t}&=&0,
\\
s_{2,t}&=& \displaystyle{\frac{3\,s_1s_2+2\,s_3s_4}{2\,s_3}},
\\
s_{3,t}&=&s_1,
\\
s_{4,t} &=&\displaystyle{
-\frac{3\,s_1\,(s_1s_2+2\,s_3s_4)}{4\,s_3^2}.}
\end{array}
\right.
\]
The general solution  of this system
\[
s_1=a_1,
\qquad
s_2 = a_4+a_3\,t,
\quad
s_3 =a_2+a_1\,t,
\qquad
s_4 = \frac{2\,a_2a_3-a_1(a_3\,t+3\,a_4)}{2\,(a_2+a_1\,t)},
\quad
a_i \in \mathbb{R},
\]
provides the four-parametric $\phi_{2,0}$--invariant solution
\[
\fl
u = \frac{a_1\,\left(\left((x+a_3t+a_4)^2+a_1t+a_2\right)^{3/2}-\left(x+a_3t+a_4\right)^3\right)}{(a_1t+a_2)^2}
-\frac{3\,a_1\,(x+a_4)+a_3\,(a_1t-2\,a_2)}{2\,(a_1t+a_2)}
\]
to equation \eqref{gHS} with $\alpha=-\frac{7}{2}$.

\section{Conclusion}
The results of the paper can be summarized as follows. We have found the Lax representation with nonremovable
spectral parameter for the generalized Hunter--Saxton equation as well as other key features of integrable
equations such as recursion operators for symmetries and cosymmetries.  We employ the recursion operators to generate the infinite-dimensional
Lie algebra of higher symmetries and then study the structure thereof, in particular we have found the basis of
this Lie algebra. We have shown that the higher symmetries from the obtained Lie algebra can be used to
construct global exact solutions for the generalized Hunter--Saxton equation. Furthermore, we have employed
recursion operators to prove existence of an infinite number of cosymmetries of higher order, which indicates
that the space of nontrivial conservation laws of higher order  is infinite-dimensional as well.

We hope that the methods used in this paper are applicable to study other properties of the generalized
Hunter--Saxton equation related to integrability such as variational symplectic and Poisson structures.
We intend to address these issues in our future work.

\section*{Acknowledgments}

This work was partially supported by the Faculty of Applied Mathematics of AGH UST statutory tasks within
subsidy of Ministry of Science and Higher Education.

I would like to express my sincere gratitude  to I.S. Kra\-{}sil${}^{\prime}$\-{}shchik for
stimulating discussions and invaluable comments.

\section*{Data availability statement}
The author confirms that the data supporting the findings of this study are available within the article.

\section*{Compliance with ethical standards}

{\bf Conflict of interest} The author declares that he has no conflict of interest.
\vskip 3 pt
\noindent
{\bf Ethical approval} The author declares that he has adhered to
the ethical standards of research execution.


\begin{thebibliography}{99}


\bibitem{Jets} %
H. Baran, M. Marvan. Jets: A software for differential calculus on jet spaces and diffieties.
Available on-line at
{\tt http://jets.math.slu.cz}

\bibitem{BealsSattingerSzmigielski2001}
R. Beals, D.H. Sattinger, J. Szmigielski.
Inverse scattering solutions of the Hunter--Saxton equation. Appl. Anal. {\bf 78} (2001), 255--269

\bibitem{Bianchi} L. Bianchi. {\it Lezioni sulla Teoria dei Gruppi Continui Finiti di Transformazioni},
E. Spoerri,  Pisa, 1918


\bibitem{BressanConstantin2005}
A. Bressan, A. Constantin. Global solutions of the Hunter--Saxton equation. SIAM J. Math. Anal. {\bf 37} (2005),
996--1026

\bibitem{Calogero1984}
F. Calogero. A solvable nonlinear wave equation. Stud. Appl. Math. {\bf 70} (1984),
189--199

\bibitem{ChernTeneblat1986}  S.S. Chern, K. Teneblat. Pseudospherical surfaces and evolution equations.
Stud. Appl. Math. {\bf 74} (1986), 55--83

\bibitem{Dryuma2001} V. Dryuma. On the Riemann and Einstein--Weyl geometry in theory of the
second order ordinary differential equations.
Bul. Acad. Sti. Rep. Moldova (Phys., Tech.), {\bf 3} (1999), 95--102;
{\tt arXiv:gr-qc/0104095}


\bibitem{DuzhinLychagin1991}
S.V. Duzhin, V.V. Lychagin. Symmetries of distributions and quad\-ra\-tu\-re of ordinary differential
equations. Acta Appl. Math.  {\bf 24} (1991), 29--57

\bibitem{Feigin1988}
B.L. Feigin. The Lie algebras $\mathfrak{gl}({\lambda})$ and cohomologies of Lie algebras of differential
operators. Russian Math. Surveys {\bf 43} (1988), 169--171

\bibitem{Golovin2004} S.V. Golovin. Group foliation of Euler equations in nonstationary rotationally symmetrical
case.  Proc. Inst. Math. of NAS of Ukraine {\bf 50} (2004), Part 1, 110--117

\bibitem{GorkaReyes2012} P. G{\'o}rka, E.G. Reyes. The modified Hunter--Saxton equation.
J. Geom. Phys. {\bf 62} (2012), 1793--1809

\bibitem{HunterSaxton1991} J.K. Hunter, R. Saxton.
Dynamics of director fields. SIAM J. Appl. Math. {\bf 51} (1991), 1498--1521

\bibitem{HunterZheng1994}
J.K. Hunter, Y.X. Zheng.
On a completely integrable nonlinear hyperbolic variational equation. Physica D {\bf 79} (1994), 361--386


\bibitem{IgoninKrasilshchik2000} %
S. Igonin, J. Krasil${}^{\prime}$shchik.
On one-parametric families of B\"acklund transformations.
In: T. Mo\-ri\-mo\-to, H. Sato, K. Yamaguchi (eds.),
{\it Lie Groups, Geometric Structures and Differential Equations --- One Hundred Years After
Sophus Lie}. Advanced Studies in Pure Mathematics, {\bf 37}, pp. 99--114.
Math. Soc. Japan, Tokyo, 2002

\bibitem{KhesinMisiolek2003} B. Khesin, G. Misio{\l}ek. Euler equations on homogeneous spaces and Virasoro orbits.
Adv. Math. {\bf 176} (2003), 116--144

\bibitem{Krasilshchik2000} %
J. Krasil${}^{\prime}$shchik.
On one-parametric families of B\"acklund transformations.
The Diffiety Institute Preprint Series. -- 2000. --  Preprint DIPS--1/2000.
Available on-line at {\tt diffiety.ac.ru}


\bibitem{KrasilshchikKersten1994}
I.S. Krasil${}^{\prime}$shchik, P.H.M. Kersten. Deformations and recursion operators
 for evolution equations // {\it Geometry in Partial Differential Equations},
Eds. A. Pr{\`{a}}staro, Th. M. Rassias,
World Sci. Publishing, River Edge, NJ, 1994, pp. 114--154


\bibitem{KrasilshchikKersten1995}
I.S. Krasil${}^{\prime}$shchik, P.H.M. Kersten. Graded differential equations and their deformations:
a computational theory for recursion operators.
Acta Appl. Math. {\bf 41} (1995), 167--191


\bibitem{KrasilshchikKersten2000}
I.S. Krasil${}^{\prime}$shchik, P.H.M. Kersten.
{\it Symmetries and Recursion Operators for Classical and Supersymmetric Differential Equations}.
Kluwer Academic Publishers, Dordrecht, 2000


\bibitem{KrasilshchikVerbovetsky2011} %
J. Krasil${}^{\prime}$shchik, A. Verbovetsky.
Geometry of jet spaces and integrable systems.
J. Geom. Phys. {\bf 61} (2011), 1633--1674

\bibitem{KrasilshchikVerbovetskyVitolo2012} %
J. Krasil${}^{\prime}$shchik, A. Verbovetsky, R. Vitolo.
A unified approach to computation of integrable struc\-tu\-res.
Acta Appl. Math. {\bf 120} (2012),  199--218

\bibitem{KrasilshchikVerbovetskyVitolo2017} 
J. Krasil${}^{\prime}$shchik, A. Verbovetsky, R. Vitolo.
{\it The Symmbolic Computation of Integrability Structures for Partial Differential Equations}.
Springer 2017

\bibitem{KrasilshchikVinogradov1984} %
I.S. Krasil${}^{\prime}$shchik, A.M. Vinogradov.
Nonlocal symmetries and the theory of coverings,
Acta Appl. Math. {\bf 2} (1984), 79--86

\bibitem{KrasilshchikVinogradov1989} %
I.S. Krasil${}^{\prime}$shchik, A.M. Vinogradov.
Nonlocal trends in the geometry of differential equations:
sym\-met\-ri\-es, conservation laws,
and B\"{a}cklund transformations.
Acta Appl. Math. {\bf 15} (1989), 161--209



\bibitem{Morozov2005} O.I. Morozov. Structure of symmetry groups via Cartan's method:
survey of four approaches. Symmetry, Integrability and Geometry: Methods and Applications,
{\bf 1} (2005), Paper 006, 14 pages

\bibitem{Olver} P.J. Olver.
{\it Applications of Lie Groups to Differential Equations}. Second Edition, Springer 1993

\bibitem{OlverRosenau1996}
P.J. Olver, Ph. Rosenau. Tri-Hamiltonian duality between solitons and solitary wave solutions having
compact support, Phys. Rev. E {\bf 53} (1996), 1900--1906

\bibitem{Ovsyannikov} L.V. Ovsiannikov. {\it Group Analysis of Differential Equations}.
Academic Press, New York, 1982

\bibitem{Pavlov2001}
M.V. Pavlov. The Calogero equation and Liouville-type equations. Theor. Math. Phys. {\bf 128} (2001),
927--932

\bibitem{PostHijligenberg1996}
G. Post, N. van den Hijligenberg.
$\mathfrak{gl}(\lambda)$ and differential operators preserving polynomials.
Acta Appl. Math.  {\bf 44} (1996), 257--268

\bibitem{Reyes2002}
E.G. Reyes. The soliton content of the Camassa--Holm and Hun\-ter--Saxton equations.
in: A.G. Nikitin, V.M. Boyko, R.O. Po\-po\-vych (Eds.), {\it Proceedings of
the Fourth International Conference on Symmetry in Nonlinear Mathematical Physics /
Proceedings of the Institute of Mathematics of the NAS of
Ukraine}, {\bf 43}, Kyiv, 2002, 201--208

\bibitem{Sasaki1979} R. Sasaki. Solton equations and pseudospherical surfaces.
Nucl. Phys. B  {\bf 154} (1979), 343--357

\bibitem{TianLiu2016}
K. Tian, Q.P. Liu. Conservation laws and symmetries of Hunter--Saxton equation: revisited.
Nonlinearity {\bf 29} (2016), 737--755

\bibitem{Tod2000}
K.P. Tod. Einstein--Weyl spaces and third order differential equations.
J. Math. Phys. {\bf 41} (2000), 5572--5581

\bibitem{VK1999} %
A.M. Vinogradov, I.S. Krasil${}^{\prime}$shchik (eds.) {\it Symmetries and Conservation Laws for Differential
Equations of Mathematical Physics} [in Russian],   Moscow: Factorial,  2005;
English transl. prev. ed.:
I.S. Krasil${}^{\prime}$shchik, A.M. Vinogradov (eds.) {\it Symmetries and Conservation Laws for Differential
Equations of Mathematical Physics}. Transl. Math. Monogr., {\bf 182}, Amer. Math. Soc., Providence, RI, 1999

\bibitem{Wang2010}
J.P. Wang. The Hunter--Saxton equation: remarkable structures of symmetries and conserved densities.
Nonlinearity {\bf 23} (2010), 2009--2028

\end{thebibliography}
\end{document}